\documentclass[12pt]{article}

\usepackage{amsfonts}
       \usepackage{amssymb}
       \usepackage{amsmath}

\def\pmx{\begin{pmatrix}}
\def\emx{\end{pmatrix}}
\def\bsq{\begin{subequations}}
\def\esq{\end{subequations}}

\def\be{\begin{eqnarray}}
\def\ee{\end{eqnarray}}
\def\bee{\begin{eqnarray*}}
\def\eee{\end{eqnarray*}}

\newtheorem{thm}{Theorem}

\newtheorem{lemma}[thm]{Lemma}

                   \def\half{{\textstyle \frac{1}{2}}}
       \def\tr{{\rm {Tr}} \, }

\def\bra{\langle}

\def\ket{\rangle}

\def\ot{\otimes}

\def\raw{\rightarrow}

\def\half{{\textstyle \frac{1}{2}}}

\def\b0{{\mathbf{\bold 0}}}

\title{New multiplicativity results for qubit maps}

\author{Christopher King and Nilufer Koldan
\\ Department of Mathematics
\\ Northeastern University
\\ Boston MA 02115
\\
}

\begin{document}

\maketitle

\begin{abstract}
Let $\Phi$ be a trace-preserving, positivity-preserving
(but not necessarily completely positive) linear map
on the algebra of complex $2 \times 2$ matrices,
and let $\Omega$ be any finite-dimensional
completely positive map. For $p=2$ and $p \geq 4$,
we prove that the maximal $p$-norm of the product map
$\Phi \ot \Omega$  is the product of the maximal $p$-norms of
$\Phi$ and $\Omega$. Restricting $\Phi$ to the class of
completely positive maps, this settles the multiplicativity question for all
qubit channels in the range of values $p \geq 4$.
\end{abstract}

\section{Introduction and statement of results}
Qubit maps provide a useful laboratory for exploring methods and conjectures
in quantum information theory. In particular they can serve as a testing ground
for approaches to the problem of
additivity of minimal entropy, and the related issues of Holevo
capacity and entanglement of formation \cite{Shor}.
In this paper we will focus on the maximal $p$-norm and consider
the question of its multiplicativity for a product map, when one of the factors
in the product is a qubit map. For values of $p$ close to one
this question is directly related to the additivity of minimal
entropy, and
hence to the circle of problems mentioned
above.

Recall first that the Schatten norm of a matrix $A$ is
defined for $p \geq 1$  as
\be\label{def:Sch}
|| A ||_{p} = \Big( \tr | A |^p \Big)^{1/p} =
\Big( \tr (A^{*} A)^{p/2} \Big)^{1/p}
\ee
Let $\Phi$ be a linear map on the matrix algebra ${\Bbb C}^{d \times d}$,
then the maximal $p$-norm of $\Phi$ is defined as
\be\label{def:max-p}
{\nu}_{p}(\Phi) = \sup_{\rho} || \Phi(\rho) ||_{p}
= \sup_{| \psi \ket} || \Phi(| \psi \ket \bra \psi |) ||_{p}
\ee
where the first $\sup$ runs over states in ${\Bbb C}^{d \times d}$,
the second $\sup$ runs over pure states (normalized vectors in ${\Bbb C}^d$),
and the second
equality follows by convexity of the $p$-norm. It is natural to
define another
norm $|| \Phi ||_{1 \raw p}$ by instead taking the $\sup$ over all matrices
$A$ satisfying $||A||_1 = 1$, and this has been considered in other work
\cite{Wat, DJKR}; however for the applications in this paper we are
interested only in the quantity defined in (\ref{def:max-p}).
In the case $d=2$ we will refer to $\Phi$ as a {\em qubit} map.

Recall that the map $\Phi$ is positivity-preserving if
$\Phi(A) \geq 0$ for every $A \geq 0$, and trace-preserving
if $\tr \Phi(A) = \tr (A)$. The map is completely positive (CP) if
in addition $\Phi \ot I_{d'}$
is positivity-preserving for every dimension $d'$. A {\em channel} is
a CP, trace-preserving map.

\medskip
Amosov and Holevo \cite{AH} conjectured that the maximal $p$-norm is
multiplicative for
products of channels, that is for any channels $\Phi$ and $\Omega$ and
for all $p \geq 1$
\be\label{AH-conj1}
{\nu}_{p}(\Phi \ot \Omega) = {\nu}_{p}(\Phi) \,\, {\nu}_{p}(\Omega)
\ee

Later Holevo and Werner \cite{WH} found a
family of $d$-dimensional channels $\Psi$
for which ${\nu}_{p}(\Psi \ot \Psi) > {\nu}_{p}(\Psi)^2$ for
$p$ sufficiently large ($p > 4.78 \dots$ for $d=3$). No such example
is known for
$d=2$, and the original conjecture (\ref{AH-conj1}) survives for the case
where at least one
of the channels $\Phi$, $\Omega$ is a qubit channel.

In our main result we prove (\ref{AH-conj1}) for the case where
$\Phi$ is a trace-preserving, positivity-preserving qubit map, where
$\Omega$ is any finite-dimensional completely positive
map, and where $p=2$ or $p \geq 4$. We do {\em not} assume that
$\Phi$ is completely positive. Indeed it is essential for our proof that we
consider the larger class of positivity-preserving but not
completely positive maps.
Previous work on entrywise positive maps \cite{KNR} has provided
other examples where
multiplicativity holds for a class of non-CP maps, in the range
$p \geq 2$.

\medskip
\begin{thm}\label{thm1}
Let $\Phi$  be a trace-preserving, positivity-preserving qubit map,
and $\Omega$
any finite-dimensional completely positive map. Then for $p=2$ and
for all $p \geq 4$,
\be\label{equal1}
{\nu}_{p}(\Phi \ot \Omega) = {\nu}_{p}(\Phi) \,\, {\nu}_{p}(\Omega)
\ee
\end{thm}

\medskip
There has been a lot of work on the additivity and multiplicativity
question for
quantum channels, and (\ref{equal1}) has been established for special classes
of qubit channels, including
the depolarizing channel \cite{BFMP, AHW, FH, A1},
unital qubit
channels \cite{KR1, Ki1}, and some classes of
non-unital qubit
channels \cite{Ki2, Ki3, ShorEBT, Fu}.
Theorem \ref{thm1} settles the question of multiplicativity for
all
qubit channels, at least in the range $p \geq 4$
(the case $p = 2$ 
was proved in \cite{Ki2}). It should be noted that
(\ref{equal1}) is false in general for positivity-preserving qubit maps
if $p < 2$, as can be seen with the example $\Phi \ot I$ where
$\Phi(\rho) = \rho^T$. We are not aware of any counterexamples to
(\ref{equal1}) for $2 < p < 4$.

\medskip
The proof of Theorem \ref{thm1} uses the following matrix inequality
which is of independent interest.

\medskip
\begin{thm}\label{thm2}
Let $A,B,C,D \in {\Bbb C}^{d \times d}$ for some $d \geq 1$.
Then for $p=2$ and for all $p \geq 4$,
\be\label{inequal1}
\bigg| \bigg| \pmx A & B \cr C & D \emx \bigg| \bigg|_{p} \leq
\bigg| \bigg| \pmx ||A||_{p} & ||B||_p \cr ||C||_p & ||D||_p \emx
\bigg| \bigg|_{p}
\ee
\end{thm}

\medskip
The inequality (\ref{inequal1}) was first derived by M. Nathanson \cite{MN}.
It had been known previously in the cases where
$A=D$ and $B=C$ \cite{BCL}, where
$\pmx A & B \cr C & D \emx$ is positive semidefinite \cite{Ki3}, and
where all matrices $A,B,C,D$ are diagonal \cite{KN}.
We conjecture that the inequality holds in the interval $2 \leq p \leq \infty$,
and that the reverse inequality holds in the interval $1 \leq p \leq  2$
(it is easy to see that equality holds at $p=2$).
Proving this conjecture would also establish the non-commutative
version of Hanner's inequality \cite{BCL}.

\medskip
The paper is organised as follows. In Section 2 we prove Theorem \ref{thm1}
for a special sub-class of qubit maps, making use of the inequality 
(\ref{inequal1}).
In Section 3 we recall a result of Gorini and Sudarshan \cite{GS} on the
classification of extreme affine maps on ${\Bbb R}^n$ which map the unit
ball into itself. Combining the Gorini-Sudarshan classification 
with 
the representation of qubit maps as affine maps on ${\Bbb R}^3$,
we derive Lemma \ref{lem2}, which implies that any trace-preserving,
positivity-preserving qubit map $\Phi$ can be expressed as a convex combination
of qubit maps from the sub-class of Section 2, all of which share the same
maximal output $p$-norm as $\Phi$. Using Lemma \ref{lem2}, we then prove
Theorem \ref{thm1} for all qubit maps. Section 4
contains the proof of Theorem \ref{thm2}, which makes use of
previously known matrix inequalities
\cite{Ki3}.

\section{Proof for special class of maps}
In this section we prove Theorem \ref{thm1} for a special class
of positivity-preserving, trace-preserving qubit maps. In order to
describe this
class we will use the representation of qubit states by points in the
Bloch sphere, and qubit maps by affine linear maps on
${\Bbb R}^3$.

A qubit state $\rho$ is represented by a point in the unit ball in ${\Bbb R}^3$
via the relation
\be
\rho = \half (I + \sum x_i \sigma_i) \mapsto x = \pmx x_1\cr x_2 \cr x_3 \emx
\ee
where $I$ is the identity matrix and $\{\sigma_1, \sigma_2, \sigma_3
\}$ are the
Pauli matrices. Positivity of $\rho$ is equivalent to
\be
\sum x_i^2 \leq 1
\ee
A trace-preserving qubit map
$\Phi$ sends the state
$\rho = \half (I + \sum x_i \sigma_i)$ to the state
$\Phi(\rho) = \half (I + \sum y_i \sigma_i)$, where $y \in {\Bbb
R}^3$ is obtained from
$x$ by applying an affine linear map, that is
\be\label{def:affine}
y = A x + v
\ee
for some real $3 \times 3$ matrix $A$, and some vector $v \in {\Bbb R}^3$.

Conjugation by a unitary matrix $U \in SU(2)$ maps $\rho$ to
$U \rho U^{*}$, and this acts on the Bloch sphere by a rotation, sending
$x \mapsto R(U) x$ for some $R(U) \in SO(3)$. If  unitary conjugations
by matrices $U,V$ are performed on the domain and range
of the map $\Phi$ respectively, then the representation
(\ref{def:affine}) is replaced by
\be\label{def:affine2}
y' = R(V) A R(U) x + R(V) v
\ee
Since the map $U \mapsto R(U)$ is onto, the singular value 
decomposition implies that
it is always possible to find
unitary matrices $U,V$ so that $R(V) A R(U)$ is diagonal
(though the diagonal entries need not be all positive).
Spectral properties of the map $\Phi$ (in particular its maximal
output $p$-norm) are invariant under unitary conjugations in its
domain and range, hence there is no loss of generality in assuming that
the matrix $A$ in (\ref{def:affine}) is diagonal.
Using the representation (\ref{def:affine}), we will say that
$\Phi$ is in {\em diagonal form} if
\be\label{diag-form}
A = \pmx \lambda_1 & 0 & 0 \cr
0 & \lambda_2 & 0 \cr
0 & 0 & \lambda_3 \emx, \quad
v = \pmx v_1 \cr v_2 \cr v_3 \emx
\ee
Note that $\Phi$ is unital
if and only if $v = 0$ in (\ref{def:affine}).
We now prove
Theorem \ref{thm1} for a special class of maps.

\medskip
\begin{lemma}\label{lemma1}
Let $\Phi$ be a positivity-preserving, trace-preserving qubit map
in diagonal form (\ref{diag-form}), and suppose that at most one
of the numbers  $(v_1, v_2, v_3)$ is nonzero.  Then
(\ref{equal1}) holds for any completely positive map $\Omega$,
for $p=2$ and for $p \geq 4$.
\end{lemma}

\medskip
\noindent{\em Proof:}
By permuting coordinates we can assume that only the third component
of $v$ can be nonzero, so that $v_1 = v_2 = 0$. The diagonal entries of
$A$ may be positive or negative. However we can change the signs of any two
diagonal entries by conjugating with a Pauli matrix, without destroying
the diagonal property and without changing the third diagonal entry;
for example
conjugating with $\sigma_3$ changes the signs of $\lambda_1$ and $\lambda_2$,
and leaves $\lambda_3$ unchanged.
Using this additional freedom we can assume that
\be\label{pos-lambda}
\lambda_1 \geq 0, \quad \lambda_2 \geq 0
\ee

Let
$\rho_{12}$ be a bipartite state on ${\Bbb C}^2 \ot {\Bbb C}^d$ for
some $d$, written
in block form
\be\label{def:rho}
\rho_{12} = \pmx X & Y \cr Y^{*} & Z \emx
\ee
Let  $\Omega$ be a completely positive map on ${\Bbb C}^d$, then
\be
(I \ot \Omega)(\rho_{12}) = \pmx A & B \cr B^{*} & C \emx
\ee
where $A = \Omega(X)$, $B = \Omega(Y)$ and $C = \Omega(Z)$.
Since $\Omega$ is completely positive, and $\rho_{12}$ is a state,
it follows that $(I \ot \Omega)(\rho_{12})$
is positive semidefinite, and hence $B = A^{1/2} R C^{1/2}$ where $R$
is a contraction. This implies in particular that for all $p \geq 1$
\be\label{2x2-pos}
|| B ||_p \leq || A ||_{p}^{1/2} \,\, || C ||_{p}^{1/2}
\ee
We will encounter the $2 \times 2$ matrices of $p$-norms
\be
\pmx || A ||_{p} & || B ||_p \cr || B ||_p & || C ||_p  \emx, \quad
\pmx || A ||_{p} & i || B ||_p \cr - i || B ||_p & || C ||_p  \emx
\ee
and we note now that (\ref{2x2-pos}) implies the positivity of these
matrices, or more generally
\be\label{pos-later}
\pmx || A ||_{p} & z || B ||_p \cr z^{*} || B ||_p & || C ||_p  \emx \geq 0
\ee
for any $z \in {\Bbb C}$ satisfying $|z| \leq 1$.

\medskip
Using the diagonal form (\ref{diag-form}) and the assumption that
$v_3$ is the only nonzero component of $v$,
we have
\be\label{mat1}
(\Phi \ot \Omega)(\rho_{12}) = \pmx c_{++} A + c_{+-} C & \lambda_1
B_1 - i \lambda_2 B_2 \cr
\lambda_1 B_1 + i \lambda_2 B_2 & c_{--} A + c_{-+} C \emx
\ee
where $B = B_1 - i B_2$ with $B_1, B_2$ hermitian, and where
\be
c_{+ \pm} = (1 + v_3 \pm \lambda_3)/2, \quad
c_{- \pm} = (1 - v_3 \pm \lambda_3)/2
\ee
Since $\Phi$ is positivity-preserving, it maps the state $\pmx 1 & 0
\cr 0 & 0 \emx$
into a positive semidefinite matrix, and this implies that
\be\label{cpos1}
c_{++} \geq 0, \quad c_{--} \geq 0
\ee
Similarly it maps the state $\pmx 0 & 0 \cr 0 & 1 \emx$ to a positive
semidefinite matrix,
hence also
\be\label{cpos2}
c_{+-} \geq 0, \quad c_{-+} \geq 0
\ee

\subsection{The case $p=2$}
Using the representation (\ref{mat1}),
\be
\tr \Big( (\Phi \ot \Omega)(\rho_{12}) \Big)^2 & = & \tr (c_{++} A +
c_{+-} C)^2 \nonumber \\
&+& 2 (\lambda_1^2 \tr B_1^2 + \lambda_2^2 \tr B_2^2)
+ \tr (c_{--} A + c_{-+} C)^2
\ee
Using the positivity of the coefficients (\ref{cpos1}), (\ref{cpos2})  and
convexity of the $2$-norm gives
\be\label{mat1.1}
\tr \Big( (\Phi \ot \Omega)(\rho_{12})\Big)^2 & \leq &  (c_{++} ||A||_2 +
c_{+-} ||C||_2)^2
\nonumber \\
& + & 2 (\lambda_1^2 \tr B_1^2 + \lambda_2^2 \tr B_2^2)
+  (c_{--} ||A||_2 + c_{-+} ||C||_2)^2
\ee
Define
\be\label{def-lambda}
\lambda = \max \{ \lambda_1, \lambda_2 \}
\ee
then it follows that
\be\label{B-ineq}
\lambda_1^2 \tr B_1^2 + \lambda_2^2 \tr B_2^2 \leq
\lambda^2 \tr B^{*} B = \lambda^2 \, || B ||_2^2
\ee
Using (\ref{B-ineq})
the right side of (\ref{mat1.1}) can be re-written as the trace
squared of a $2 \times 2$
matrix, leading to
\be\label{mat1.2}
\tr \Big( (\Phi \ot \Omega)(\rho_{12})\Big)^2 & \leq &
\tr  \pmx c_{++} ||A||_2 + c_{+-} ||C||_{2} & \lambda ||B||_{2} \cr
\lambda ||B||_2 & c_{--} ||A||_2 + c_{-+} ||C||_2 \emx^2 \nonumber \\
& = & \tr \bigg( \Phi \pmx ||A||_2 & z ||B||_{2} \cr
z^{*} ||B||_2 & ||C||_2 \emx \bigg)^2
\ee
where
\be\label{def-z}
z = \begin{cases}1 & \mbox{if} \,\, \lambda = \lambda_1 \\
i & \mbox{if} \,\, \lambda = \lambda_2 \end{cases}
\ee
As noted in (\ref{pos-later}) the matrix $\pmx ||A||_2 & z ||B||_{2} \cr
z^{*} ||B||_2 & ||C||_2 \emx$ is positive semidefinite, hence by definition
of the maximal $2$-norm we get
\be
||(\Phi \ot \Omega)(\rho_{12})||_{2} & \leq &
\bigg| \bigg| \Phi \pmx ||A||_2 & z ||B||_{2} \cr
z^{*} ||B||_2 & ||C||_2 \emx \bigg| \bigg|_{2} \nonumber \\
& \leq & \nu_{2}(\Phi) \,\, \tr \pmx ||A||_2 & z ||B||_{2} \cr
z^{*} ||B||_2 & ||C||_2 \emx \nonumber \\
& =& \nu_{2}(\Phi) \,\, (||A||_2 + ||C||_2)
\ee
Since $A = \Omega(X)$ and $C = \Omega(Z)$, this yields
\be
||(\Phi \ot \Omega)(\rho_{12})||_{2} & \leq &
\nu_{2}(\Phi) \, \nu_{2}(\Omega) \, (\tr X + \tr Z) \nonumber \\
& = & \nu_{2}(\Phi) \, \nu_{2}(\Omega)
\ee
since $\tr X + \tr Z = \tr \rho_{12} = 1$. Since this holds for any state
$\rho_{12}$ we deduce that
\be
\nu_{2}(\Phi \ot \Omega) \leq \nu_{2}(\Phi) \, \nu_{2}(\Omega)
\ee
The inequality in the reverse direction follows by restriction to product
states, hence this completes the proof for the
case $p=2$.

\subsection{The case $p \geq 4$}
We apply Theorem \ref{thm2} to (\ref{mat1})
    to conclude that for
$p \geq 4$,
\be\label{mat2}
||(\Phi \ot \Omega)(\rho_{12})||_{p} \leq
\bigg| \bigg| \pmx ||c_{++} A + c_{+-} C||_{p} & ||\lambda_1 B_1 - i
\lambda_2 B_2||_p \cr
||\lambda_1 B_1 + i \lambda_2 B_2||_p & ||c_{--} A + c_{-+} C||_p
\emx \bigg| \bigg|_{p}
\ee
Define the $2 \times 2$ real symmetric matrix
\be\label{mat3}
M = \pmx ||c_{++} A + c_{+-} C||_{p} & ||\lambda_1 B_1 - i \lambda_2
B_2||_p \cr
||\lambda_1 B_1 + i \lambda_2 B_2||_p & ||c_{--} A + c_{-+} C||_p \emx,
\ee
so that (\ref{mat2}) can be written
\be\label{mat2.5}
||(\Phi \ot \Omega)(\rho_{12})||_{p} \leq
|| M ||_{p}
\ee

The positivity results (\ref{cpos1}) and (\ref{cpos2})
imply that
\be\label{cpos3}
||c_{++} A + c_{+-} C||_{p} & \leq & c_{++} ||A||_p + c_{+-}
||C||_{p}, \nonumber \\
||c_{--} A + c_{-+} C||_p & \leq & c_{--} ||A||_{p} + c_{-+} ||C||_p
\ee
Furthermore, recall (\ref{def-lambda}) and
suppose first that $\lambda = \lambda_1$, so that
$\lambda_1 - \lambda_2 \geq 0$. Then
\be\label{bound-B}
||\lambda_1 B_1 - i \lambda_2 B_2||_p & = &
||(\lambda_1 - \lambda_2) B_1 + \lambda_2 B||_p \nonumber \\
& \leq & (\lambda_1 - \lambda_2) ||B_1 ||_p + \lambda_2 ||B||_p \nonumber \\
& \leq & \lambda || B ||_p
\ee
where in the last inequality we used $||B_1||_p = \half ||B +
B^{*}||_p \leq ||B||_p$.
A similar argument leads to the same conclusion if $\lambda = \lambda_2$.

We would like to replace the entries of $M$ with the bounds on the right
sides of (\ref{cpos3}) and (\ref{bound-B}),
and argue that $|| M ||_{p}$ must increase under this substitution.
However the matrix $M$ may not be positive semidefinite
(since $\Phi$ is not necessarily completely positive) so this is not
immediately obvious.
To see that it does in fact increase, let $p = 2 q$ so that
\be
|| M ||_{p} = \Big( || M^2 ||_{q} \Big)^{1/2}
\ee
Then the matrix $M^2 = M^{*} M$ is positive semidefinite with
positive
entries, and it is easy to see that this implies $|| M^2 ||_{q}$ is an
increasing
function of the entries of $M^2$. Since $M$ is also entrywise
positive,  the entries of $M^2$ are increasing functions of the entries of $M$,
and therefore so is $|| M^2 ||_{q}$. Therefore $|| M ||_{p}$ increases when
the bounds (\ref{cpos3}), (\ref{bound-B}) are inserted in the right
side of (\ref{mat2.5}),
and we get
\be\label{mat4}
||(\Phi \ot \Omega)(\rho_{12})||_{p} & \leq  &
\bigg| \bigg| \pmx c_{++} ||A||_p + c_{+-} ||C||_{p} & \lambda ||B||_p \cr
\lambda ||B||_p & c_{--} ||A||_{p} + c_{-+} ||C||_p \emx \bigg|
\bigg|_{p}
\ee
Now we note that the right side of
(\ref{mat4}) is unchanged if the upper-right entry $\lambda ||B||_p$
is replaced by $z \lambda ||B||_p$ and the lower left
entry by  $z^{*} \lambda ||B||_p$ for any $|z| = 1$.
Hence using the notation (\ref{def-z}) again, (\ref{mat4}) implies
\be\label{mat4.5}
||(\Phi \ot \Omega)(\rho_{12})||_{p} & \leq & \bigg| \bigg| \Phi \pmx
||A||_p &  z ||B||_p \cr
z^{*} ||B||_p &  ||C||_p \emx \bigg| \bigg|_{p}
\ee
We now repeat the arguments used above in the case $p=2$, to conclude that
\be\label{mat5}
||(\Phi \ot \Omega)(\rho_{12})||_{p} & \leq &
\nu_{p}(\Phi) \, (||A||_p + ||C||_p ) \nonumber \\
& \leq & \nu_{p}(\Phi) \, \nu_{p}(\Omega) \, (\tr X + \tr Z) \nonumber \\
& = & \nu_{p}(\Phi) \, \nu_{p}(\Omega)
\ee
Since this holds for any state $\rho_{12}$ we again deduce
\be
\nu_{p}(\Phi \ot \Omega) \leq
\nu_{p}(\Phi) \, \nu_{p}(\Omega)
\ee
and this completes the proof for the case $p \geq 4$.

\section{Reduction to special form}
In this section we will show that the general case of Theorem \ref{thm1}
follows from Lemma \ref{lemma1}. Recall that
a trace-preserving, 
positivity-preserving qubit map $\Phi$ is represented
by an affine linear map on ${\Bbb R}^3$ as in   (\ref{def:affine}),
sending the Bloch sphere (the closed unit ball in ${\Bbb R}^3$)
into 
an ellipsoid.  We will refer to the latter as the {\em
image ellipsoid} of $\Phi$.

\medskip
For a positivity-preserving, trace-preserving qubit map $\Phi$,
the minimal output entropy and maximal output $p$-norm
are all achieved on the same input state. That is, there is a pure state
$| \psi \ket$ such that for all $p \geq 1$
\be\label{def:p-max1}
{\nu}_{p}(\Phi) = \sup_{\rho} ||\Phi(\rho)||_{p} =
||\Phi(| \psi \ket \bra \psi|)||_{p}
\ee
Define the function
\be\label{def:h}
h_{p}(r) = \bigg(\Big({1+r \over 2}\Big)^p + \Big({1 -r \over
2}\Big)^p \bigg)^{1/p}
\ee
The spectrum of $\Phi(| \psi \ket \bra \psi |)$ is $\{(1 \pm r)/2\}$, for some
$0 \leq r \leq 1$. Accordingly the value of (\ref{def:p-max1}) is
\be\label{eval:p-max}
{\nu}_{p}(\Phi) =  h_{p}(r)
\ee

We will denote by ${\cal C}_r$ the set of all positivity-preserving,
trace-preserving
qubit maps whose maximal output $p$-norm is at most $h_{p}(r)$, that is
\be\label{def:calC}
{\cal C}_r = \{ \Phi \, : \, {\nu}_{p}(\Phi) \leq  h_{p}(r) \}
\ee
Note that ${\cal C}_r$ does not depend on $p$. Geometrically,
${\cal C}_r$ consists of the positivity-preserving qubit maps for
which the image ellipsoid
lies inside the sphere of radius $r$ centered at the origin.

It is clear that ${\cal C}_r$ is a convex set. The next result shows
that the extreme
points of ${\cal C}_r$ have a simple form. Recall the definition 
(\ref{diag-form}) 
of the diagonal form of a qubit map.

\medskip
\begin{lemma}\label{lem2}
Let $\Phi$ be an extreme point in ${\cal C}_r$, represented in 
diagonal form by the
affine map $x \mapsto A x + v$ on ${\Bbb R}^3$. 
Then at most one of the components of $v$ is nonzero.
\end{lemma}

\medskip
Lemma \ref{lem2} is a consequence of the following 
Theorem of Gorini 
and Sudarshan \cite{GS},
which classifies all extreme affine maps of 
${\Bbb R}^n$ sending the closed
unit ball into itself.

\medskip
\begin{thm}\label{thmGS}[Gorini-Sudarshan]
Let $D_n$ be 
the set of affine maps of ${\Bbb R}^n$ which send
the closed unit 
ball into itself. Denote by $(B, w)$ the map
$x \mapsto B x + w$, 
where $w \in {\Bbb R}^n$ and $B \in {\Bbb R}^{n \times n}$.
If $(B, 
w)$ is an extreme point in $D_n$, then there are orthogonal 
matrices
$Q_1, Q_2 \in O(n)$, and real numbers $0 \leq \kappa \leq 
1$, $0 < \delta \leq 1$
such that
\be\label{GS1}
Q_1 w = (0, \cdots, 
0, \delta(1 - \kappa^2)), \quad
Q_1 B Q_2 = {\rm Diag}(m,  \cdots, m, 
\kappa m)
\ee
where $m = \sqrt{1 + \kappa^2 \delta^2 - \delta^2}$ and 
where
${\rm Diag}(d_1, d_2, \dots)$ denotes the diagonal matrix with 
entries
$d_1, d_2, \dots$.
\end{thm}

\medskip
To derive Lemma 
\ref{lem2} from Theorem \ref{thmGS}, we identify ${\cal C}_r$
with 
the set of scaled affine maps $r D_3 = \{ (rB, rw) \,:\, (B,w) \in 
D_3 \}$.
Hence every extreme map $\Phi$ in ${\cal C}_r$ corresponds 
to
an affine map $(rB,rw)$ where
$(B,w)$ satisfies (\ref{GS1}). 
Furthermore the matrix $Q_1$ in (\ref{GS1}) is in $O(3)$, and 
hence
either $Q_1 \in SO(3)$ or $- Q_1 \in SO(3)$; similarly for 
$Q_2$.
Since every rotation in $SO(3)$ can be implemented by a 
unitary conjugation
in $SU(2)$ (see the discussion leading up to 
(\ref{def:affine2})), this shows that $\Phi$ can be written in 
diagonal form 
with
\be\label{diag-form2}
A = \pmx \pm r m & 0 & 0 \cr
0 & \pm r m & 0 \cr
0 & 0 & \pm r \kappa m \emx, \quad
v = \pmx 0 \cr 0 \cr \pm r \delta(1 - \kappa^2) \emx,
\ee
and this proves  Lemma \ref{lem2}.

\medskip
In the remainder of this section
we will  show that
Theorem \ref{thm1} follows from Lemma \ref{lemma1} and Lemma \ref{lem2}.
Accordingly,
suppose that $\Phi$ is a trace-preserving, positivity-preserving
qubit map satisfying (\ref{eval:p-max}) for some
$0 \leq r \leq 1$, so that
\be\label{def:p-max}
{\nu}_{p}(\Phi) = h_{p}(r)
\ee
Then it is sufficient to show that for any completely positive map $\Omega$,
\be\label{suff1}
\nu_{p}(\Phi \ot \Omega) \leq
h_{p}(r) \,\, \nu_{p}(\Omega)
\ee

Now ${\cal C}_r$ is a closed bounded convex subset of ${\Bbb R}^{12}$
(since the matrix $A$ and vector $v$ together have $12$ entries),
hence by Caratheodory's Theorem any element of ${\cal C}_r$
can be written as a convex combination of at most $13$ of its extreme points.
The map $\Phi$ is in ${\cal C}_r$, hence there are extreme maps
$\{{\Phi}_i\} \in {\cal C}_r$ such that
\be\label{sum1}
\Phi = \sum_{i} a_i {\Phi}_i
\ee
where $a_i \geq 0$ and $\sum a_i = 1$. Since $\{\Phi_i\}$ are
in 
${\cal C}_r$ we also have
\be
{\nu}_{p}(\Phi_i) \leq h_{p}(r)
\ee
Furthermore, combining Lemma \ref{lem2} and Lemma \ref{lemma1},
we deduce that
\be\label{ext-case}
{\nu}_{p}(\Phi_i \ot \Omega) = {\nu}_{p}(\Phi_i) \,\, \nu_{p}(\Omega) \leq
h_{p}(r) \, \nu_{p}(\Omega)
\ee
for all $i$. By convexity of the $p$-norm it follows from (\ref{sum1}) and
(\ref{ext-case}) that
\be\label{convex}
{\nu}_{p}(\Phi \ot \Omega) & \leq & \sum_{i} a_i \,
{\nu}_{p}({\Phi}_i \ot \Omega) \nonumber \\
& = & \sum_{i} a_i \, {\nu}_{p}({\Phi}_i) \,\, {\nu}_{p}(\Omega) \nonumber \\
& \leq & h_{p}(r) \, \nu_{p}(\Omega)
\ee
and this proves (\ref{suff1}).

\section{Proof of Theorem \ref{thm2}}
Let $p = 2 q$ and define
\be
M
= \pmx A & B \cr C & D \emx
\ee
Then $M^{*} M$ is positive
semidefinite, and we write it in block form as
\be
M^{*} M = \pmx
M_{11} & M_{12} \cr M_{21} & M_{22} \emx
\ee
Now we apply the result
of Theorem 1(b) from \cite{Ki3} to the matrix
$M^{*} M$ to deduce that
\be\label{q-bound}
|| M^{*} M ||_{q} \leq
\bigg| \bigg| \pmx ||M_{11}||_{q} & ||M_{12}||_q
\cr ||M_{21}||_q &
||M_{22}||_q \emx \bigg| \bigg|_{q}
\ee
for all $q \geq 2$. Furthermore
$M_{11} = A^{*} A + C^{*} C$,
hence
\be\label{m11}
||M_{11}||_q \leq ||A^{*} A||_q + ||C^{*} C||_q
=
|| A ||_{p}^2 + || C
||_{p}^2
\ee
Similarly
\be\label{m12}
||M_{12}||_q = || M_{21} ||_q
\leq || A ||_{p} \, || B ||_p + || C ||_{p} \, || D
||_p
\ee
and
\be\label{m22}
||M_{22}||_q \leq || B ||_{p}^2 + || D
||_{p}^2
\ee
For a positive semidefinite $2 \times 2$ matrix the
$q$-norm is an increasing
function of the entries. Hence combining (\ref{q-bound})
with
(\ref{m11}), (\ref{m12}), (\ref{m22}) gives
\be
|| M^{*} M ||_{q}
\leq || m^{*} m ||_q
\ee
where
\be
m = \pmx ||A||_p & ||B||_p \cr ||C||_p & ||D||_p
\emx
\ee
Taking a square root of both sides gives
\be
|| M ||_p \leq
|| m ||_p
\ee
which is the stated result.

\bigskip
{\bf Acknowledgements} This research
was supported in part by National Science Foundation Grant
DMS-0400426. The authors are grateful to Michael Nathanson
for discussions about his earlier work on the inequalities in Theorem 
2.
The authors are also grateful to the referee for pointing them 
toward
reference \cite{GS}.

{~~}

\end{document}